\documentclass[twocolumn,preprintnumbers,amsmath,amssymbm,prl]{revtex4}
\usepackage{epsfig}
\usepackage{graphicx}

\begin{document}
\title{Lower bound on the value of the fine-structure constant}
\author{Shahar Hod}
\affiliation{The Ruppin Academic Center, Emeq Hefer 40250, Israel}
\affiliation{ } \affiliation{The Hadassah Institute, Jerusalem
91010, Israel}
\date{\today}

\begin{abstract}
\ \ \ Recently we have proposed the existence of a universal
relation between the maximal electric charge and total mass of any
weakly self-gravitating object: $Z\leq Z^*={\alpha}^{-1/3}A^{2/3}$,
where $Z$ is the number of protons, $A$ is the total baryon (mass)
number, and $\alpha=e^2/\hbar c$ is the fine-structure constant.
Motivated by this novel bound, we explore the $(Z,A)$-relation of
atomic nuclei as deduced from the Weizs\"acker semi-empirical mass
formula. It is shown that {\it all} nuclei, including the
meta-stable maximally charged ones, conform to the upper bound.
Moreover, we suggest that the new charge-mass bound places an
interesting constraint on the value of the fine-structure constant:
$\alpha\gtrsim 1/323$.
\end{abstract}
\bigskip
\maketitle


The weak cosmic censorship conjecture (WCCC), introduced by Penrose
forty years ago \cite{HawPen,Pen}, is one of the corner stones of
general relativity. This principle asserts that spacetime
singularities that arise in gravitational collapse are always hidden
inside of black holes. The elimination of a black-hole horizon is
ruled out by this hypothesis because that would expose naked
singularities to distant observers.

Arguing from the cosmic censorship principle, we have proposed
\cite{Hodbound} the existence of a universal bound on the charge $q$
of any weakly self-gravitating object of total energy $\mu$:
$q\leq\mu^{2/3}E^{-1/3}_c$, where $E_c$ is the critical electric
field for pair-production \cite{Emax1}. For charged objects with
nuclear matter density the upper bound corresponds to
\begin{equation}\label{Eq1}
Z\leq Z^*={\alpha}^{-1/3}A^{2/3}\  ,
\end{equation}
where $Z$ and $A$ are the number of protons and the total baryon
number, respectively, and $\alpha\equiv e^2/\hbar c$ is the
fine-structure constant. [We shall henceforth use natural units in
which $c=1$.]

This bound was inferred from the requirement that the WCCC be
respected in a gedanken experiment in which a charged object falls
into a charged black hole. The integrity of the black-hole horizon
is respected provided $Z$ is bounded as in Eq. (\ref{Eq1}). This
relation limits the charges of objects such as atomic nuclei and
quark nuggets \cite{Hodbound,Mad1}. The intriguing feature of our
derivation \cite{Hodbound} is that it uses a principle whose very
meaning stems from gravitation (the cosmic censorship principle) to
derive a universal bound which has nothing to do with gravitation
[written out fully, the bound (\ref{Eq1}) would involve $\hbar$ and
$c$, but not $G$]. This provides a striking illustration of the
unity of physics.

It is of considerable interest to check the validity of the new
charge-mass bound (\ref{Eq1}). For instance, Lead $_{82}^{208}{Pb}$,
the largest known completely stable nucleus satisfies the relation
$Z/A^{2/3}\simeq 2.33<{\alpha}^{-1/3}$. Thus, this nucleus conforms
to the upper bound (\ref{Eq1}). The largest known artificially made
nucleus contains $Z=118$ protons and a total number of $A=294$
nucleons \cite{Ogan}. This nucleus satisfies the relation
$Z/A^{2/3}\simeq 2.67<{\alpha}^{-1/3}$. Thus, one finds that this
nucleus also respects the upper bound (\ref{Eq1}) \cite{Hodbound}.

Even heavier meta-stable nuclei are expected to be produced in the
forthcoming years using accelerator production techniques. In fact,
some calculations suggest that nuclei of $A\sim 300$ to $476$ may
have very long lifetimes \cite{Nato}. Could these highly charged
nuclei be able to threaten the validity of the cosmic censorship
conjecture by violating the $(Z,A)$-bound (\ref{Eq1})? In order to
address this question, we shall investigate the charge-mass relation
of atomic nuclei as deduced from the well-known semi-empirical mass
formula \cite{Rohlf,Segre,Cook,Goo1,Kir}.

Consider an atomic nucleus composed of $Z$ protons and $N$ neutrons.
The total baryon number is given by $A=Z+N$. The charge and mass of
a nucleus are given by
\begin{equation}\label{Eq2}
q=Z|e|\ \ \ ; \ \ \ \mu=Zm_p+Nm_n-{\cal E}_B\simeq Am_p\  ,
\end{equation}
where ${\cal E}_B$ is the binding energy of the nucleus, which is
typically much smaller than its mass. The binding energy ${\cal
E}_B$ of a nucleus (that is, the difference between its mass and the
sum of the masses of its individual constituents) is well
approximated by the semi-empirical mass formula, also known as
Weizs\"acker's formula \cite{Rohlf,Segre,Cook,Goo1,Kir}:
\begin{eqnarray}\label{Eq3}
{\cal E}_B(A,Z)=a_V A-a_S A^{2/3}-a_C
{{Z^2}\over{A^{1/3}}}\nonumber\\-a_A{{{(A-2Z)}^2}\over{A}}\ .
\end{eqnarray}

This well-known formula is based on the liquid drop model which
treats the nucleus as a drop of incompressible nuclear fluid
composed of protons and neutrons. The four terms on the r.h.s of Eq.
(\ref{Eq3}) correspond to the cohesive binding of all the nucleons
by the strong nuclear force, a surface energy term (which represents
the fact that surface nucleons are less tightly bound as compared to
bulk nucleons), the electrostatic mutual repulsion of the protons,
and an asymmetry term (which represents the fact that protons and
neutrons occupy independent quantum momentum states)
\cite{Rohlf,Segre,Cook,Goo1,Kir}. The coefficients in the
semi-empirical mass formula are calculated by fitting to
experimentally measured masses of nuclei \cite{Kir}:
\begin{eqnarray}\label{Eq4}
&a_V=15.36(3)\ \ \ ;\ \ a_S=16.42(8)\ \ ;\nonumber\\ &a_C=0.691(2)\
\ ;\ \ a_A=22.53(7)\  .&
\end{eqnarray}

To a rough approximation, the nucleus can be considered a sphere of
uniform charge density. The potential energy of such a charge
distribution is given by ${\cal E}_p=3Q^2/5R$. It is well-known that
the radii of atomic nuclei are well approximated by the empirical
relation $R(A)=1.219\times A^{1/3}\ {\text{fm}}$ \cite{Kir}. This is
a direct consequence of the fact that the size of an individual
nucleon is roughly given by its Compton length. Thus, one can write
$R(A)\simeq A^{1/3}\times{{\xi\hbar}\over{m_p}}\ {\text{fm}}$, where
$m_p$ is the proton's mass and $\xi$ is a constant of order unity
(empirically one finds $\xi\simeq 5.788$). Substituting $R(A)$ and
$Q=Z|e|$ into the expression of the potential energy ${\cal E}_p$,
one finds
\begin{equation}\label{Eq5}
a_C\simeq {{3m_p\alpha}\over{5\xi}}\equiv c\alpha\  ,
\end{equation}
where $c\equiv 3m_p/5\xi\simeq 97.38$. For $\alpha\simeq 1/137.036$
\cite{Han} one finds $a_C\simeq 0.711$. This estimated value of
$a_C$ is astonishingly close to the empirically measured one (less
than $3\%$ difference), see Eq. (\ref{Eq4}). Below we shall come
back to this observation.

We shall first consider nuclei with the {\it largest} possible
electric charge, $Z_{\text{max}}(A)$, for a given value of the
baryon number $A$. These nuclei pose the greatest challenge to the
charge-mass bound (\ref{Eq1}). A nucleus may in principle be
produced (and live even for a short duration of time) if it has a
positive binding energy. The maximal values $Z_{\text{max}}(A)$ are
determined from Eq. (\ref{Eq3}) with the requirement ${\cal
E}_B(A,Z_{\text{max}})\geq0$ [with ${\cal
E}_B(A,Z_{\text{max}}+1)<0$]. It should be emphasized that, although
such hypothetical nuclei are characterized by positive binding
energy (that is, their masses are {\it smaller} than the sum of the
masses of their individual constituents), they are expected to be
short-lived. This is due to the fact that their binding energies are
smaller than the corresponding binding energies of the stable
nuclei, see Eq. (\ref{Eq7}) below. Thus, we do not expect to find
such nuclei in nature. Nevertheless, such nuclei could in principle
be produced artificially (and live for a short duration of time),
and it is therefore of interest to study these nuclei in the context
of the new charge-mass bound (\ref{Eq1}).

Noting that the second and fourth terms in the mass formula
(\ref{Eq3}) are negative, one realizes that the binding energy
${\cal E}_B(A,Z)$ will become negative once the Coulomb energy
overcomes the volume energy \cite{NoteAv}. Thus, one may obtain a
simple upper bound on the value of $Z_{\text{max}}(A)$:
\begin{equation}\label{Eq6}
Z_{\text{max}}(A)<\Big({{a_V}\over{a_C}}\Big)^{1/2}A^{2/3}\  .
\end{equation}
Substituting the experimentally measured values of the coefficients
$a_V$ and $a_C$ \cite{Kir}, one finds
$Z_{\text{max}}(A)<4.71A^{2/3}$. We therefore obtain the series of
inequalities $Z(A)/Z^*(A)\leq
Z_{\text{max}}(A)/Z^*(A)<4.71/\alpha^{-1/3}<0.91<1$. This implies
that all atomic nuclei, including the meta-stable {\it maximally}
charged ones (with the maximally allowed electric charge according
to the semi-empirical mass formula. Most of these nuclei are yet to
be produced artificially) conform to the new $Z-A$ bound
(\ref{Eq1}).

\begin{figure}[h]
  \begin{center}
    \epsfxsize=8.5cm \epsffile{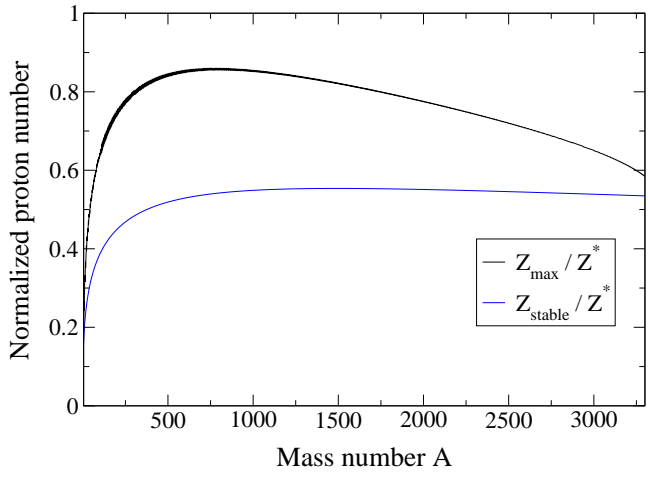}
  \end{center}
  \caption{The ratios $Z_{\text{max}}(A)/Z^*(A)$ and $Z_{\text{stable}}(A)/Z^*(A)$
as a function of the mass number $A$. Here $Z_{\text{max}}(A)$,
$Z_{\text{stable}}(A)$, and $Z^*(A)$ are the maximally allowed
number of protons in a nucleus of mass number $A$ according to the
Weizs\"acker semi-empirical mass formula (\ref{Eq3}), the number of
protons of the most stable nucleus of atomic mass $A$, and the
maximally allowed number of protons for a given mass number $A$
according to the novel upper bound (\ref{Eq1}), respectively. One
realizes that these ratios are smaller than $1$ for {\it all} nuclei
(including hypothetical heavy nuclei with positive binding energy
which are yet to be produced), with a maximal value of $\simeq
0.85$.}
  \label{Fig1}
\end{figure}

In figure \ref{Fig1} we depict the actual ratio
$Z_{\text{max}}(A)/Z^*(A)$ as a function of the mass number $A$,
where $Z_{\text{max}}(A)$ is calculated from the full expression of
the Weizs\"acker semi-empirical mass formula (\ref{Eq3}). One indeed
finds $Z_{\text{max}}/Z^*<1$ for {\it all} nuclei, with a maximal
value of $\simeq 0.85$.

In figure \ref{Fig1} we also depict the ratio
$Z_{\text{stable}}(A)/Z^*(A)$ as a function of the mass number $A$,
where $Z_{\text{stable}}(A)$ is the number of protons of the most
stable nucleus of mass number $A$. Here $Z_{\text{stable}}(A)$ is
obtained by maximizing the binding energy ${\cal E}_B(A,Z)$ with
respect to $Z$. This yields \cite{Hodbound}
\begin{equation}\label{Eq7}
Z_{\text{stable}}(A)={A\over 2}{1\over{1+{{a_C}\over{4a_A}}
A^{2/3}}}\  .
\end{equation}
(For light nuclei this expression reduces to the canonical relation
$Z=A/2$.) One finds that the ratio $Z_{\text{stable}}/Z^*$ has a
maximal value of $\simeq 0.56$ \cite{Hodbound}.

The dimensionless fine-structure constant $\alpha$ is one of the
fundamental parameters of the standard model of particle physics. It
has puzzled many scientists since its introduction by Sommerfeld
\cite{Sommerfeld} almost a century ago. The question of how to
derive the numerical value of $\alpha$ from some underlying theory
has been one of the most important open questions in modern physics
\cite{Dirac,Eddington,Das,Boyer,Wyler,Rosen,Chew,Ross,Dav}. As we
shall now argue, one may deduce a bound on the value of the
fine-structure constant $\alpha$ from the novel charge-mass bound
(\ref{Eq1}). To that end, let us assume that $\alpha$ is a free
parameter in Eqs. (\ref{Eq1}) and (\ref{Eq5}).

The requirement $Z_{\text{stable}}(A)\leq Z^*(A)$ yields the
quadratic equation
\begin{equation}\label{Eq8}
{{a_C}\over{2a_A}}A^{2/3}-{\alpha}^{1/3}A^{1/3}+2\geq 0\  .
\end{equation}
The inequality (\ref{Eq8}) would be satisfied for {\it all} $A$
values provided the discriminant is non-positive \cite{Notealpha0}:
\begin{equation}\label{Eq9}
\alpha^{2/3}-{{4a_C}\over{a_A}}\leq 0\  .
\end{equation}
Substituting $a_C=c\alpha$ from Eq. (\ref{Eq5}), one obtains the
inequality:
\begin{equation}\label{Eq10}
\alpha\geq \Big({{a_A}\over{4c}}\Big)^3\  .
\end{equation}
Substituting the experimentally measured values of the coefficients
$a_A$ and $c$ \cite{Kir}, one finds a lower bound on the value of
the fine structure constant:
\begin{equation}\label{Eq11}
\alpha\gtrsim {{1}\over{5161.7}}\  .
\end{equation}

We shall now consider atomic nuclei with the largest possible
electric charge for a given mass number $A$.
In order to avoid a violation of the bound (\ref{Eq1}) (which would
ultimately lead to a violation of the WCCC, see \cite{Hodbound}), we
must demand that any collection of nucleons which seems to violate
(\ref{Eq1}) is actually unstable. More explicitly, we must demand
that any collection of $Z$ protons and $A-Z$ neutrons with
$Z>{\alpha}^{-1/3}A^{2/3}$ is characterized by a negative binding
energy. Substituting $Z^*={\alpha}^{-1/3}A^{2/3}$ from the bound
(\ref{Eq1}) into (\ref{Eq3}) and demanding that ${\cal
E}_B(A,Z^*+1)<0$, one obtains the quadratic equation
\begin{equation}\label{Eq12}
(a_V-{{a_C}\over{\alpha^{2/3}}}-a_A)A^{2/3}+({{4a_A}\over{\alpha^{1/3}}}-a_S)A^{1/3}-
{{4a_A}\over{\alpha^{2/3}}}\leq 0\  .
\end{equation}
This inequality would be respected by {\it all} nuclei (by all $A$
values) provided the discriminant is non-positive \cite{Notealpha1}:
\begin{equation}\label{Eq13}
({{4a_A}\over{\alpha^{1/3}}}-a_S)^2+{{16a_A}\over{\alpha^{2/3}}}
(a_V-{{a_C}\over{\alpha^{2/3}}}-a_A)\leq 0\  .
\end{equation}
Substituting $a_C=c\alpha$ from Eq. (\ref{Eq5}), the inequality
(\ref{Eq13}) can be written as a quadratic equation for
$\alpha^{1/3}$:
\begin{equation}\label{Eq14}
a^2_S\alpha^{2/3}-8a_A(2c+a_S)\alpha^{1/3}+16a_Aa_V\leq 0\ .
\end{equation}
This yields the lower bound
\begin{equation}\label{Eq15}
\alpha^{1/3}\geq
{{4a_A(2c+a_S)-4\sqrt{a^2_A(2c+a_S)^2-a^2_Sa_Aa_V}}\over{a^2_S}}\ .
\end{equation}
Substituting the experimentally measured values of the coefficients
$a_V,\ a_S,\ a_A$, and $c$ \cite{Kir}, one finds the lower bound
\cite{Note1}
\begin{equation}\label{Eq16}
\alpha\gtrsim {{1}\over{323.6}}\  .
\end{equation}
This bound is necessary for the validity of the charge-mass relation
(\ref{Eq1}) \cite{Notealpha2}. Remarkably, the bound (\ref{Eq16}) on
the value of the fine-structure constant is of the {\it same} order
of magnitude as the experimentally measured value \cite{Han}.

\bigskip
\noindent {\bf ACKNOWLEDGMENTS}

This research is supported by the Meltzer Science Foundation. I
thank Oded Hod, Yael Oren and Arbel M. Ongo for helpful discussions.


\newpage

\input{epsf}


\begin{thebibliography}{99}

\bibitem{HawPen} S. W. Hawking and R. Penrose, Proc. R. Soc. London
A {\bf 314}, 529 (1970).

\bibitem{Pen} R. Penrose, Riv. Nuovo Cimento {bf 1}, 252 (1969); in
{\it General Relativity, an Einstein Centenary Survey}, edited by S.
W. Hawking and W. Israel (Cambridge University Press, Cambridge,
England, 1979).

\bibitem{Hodbound} S. Hod, Phys. Lett. B {\bf 693}, 339 (2010) [arXiv:1009.3695].

\bibitem{Emax1} F. Sauter, Z. Phys. {\bf 69}, 742 (1931); W. Heisenberg and H. Euler, Z. Phys. {\bf 98}, 714 (1936);
J. Schwinger, Phys. Rev. {\bf 82}, 664 (1951).

\bibitem{Mad1} J. Madsen, Phys. Rev. Lett. {bf 100}, 151102 (2008).

\bibitem{Ogan} Y. T. Oganessian et. al., Phys. Rev. C {\bf 74}, 044602 (2006).

\bibitem{Nato} See J. B. Natowitz, Physics {\bf 1}, 12 (2008) and references therein.

\bibitem{Rohlf} J. W. Rohlf, {\it Modern Physics from a to Z0} (Wiley,
1994).

\bibitem{Segre} E. Segr\'e, {\it Nuclei and Particles} (Second
Edition, W. A. Benjamin, 1977).

\bibitem{Cook} N. Cook, {\it Models of the Atomic Nucleus} (Springer Verlag,
2006).

\bibitem{Goo1} See also http://en.wikipedia.org/wiki/Semi-empirical\_mass\_formula.

\bibitem{Kir} M. W. Kirson, Nucl. Phys. A {\bf 798}, 29 (2008).

\bibitem{NoteAv} A similar argument was raised by A. Gal (private
communication).

\bibitem{Han} D. Hanneke, S. Fogwell, and G. Gabrielse, Phys. Rev. Lett. {\bf 100}, 120801 (2008).

\bibitem{Sommerfeld} A. Sommerfeld, {\it Atombau und
Spectrallinien}, (Friedr. Vieweg \& Sohn, Brounschweig, 1951).

\bibitem{Dirac} P. A. M. Dirac, Proc. R. Soc. (London) A{\bf 133},
60 (1931).

\bibitem{Eddington} A. S. Eddington, {\it Fundamental Theory} (Cambridge University Press, Cambridge
1946).

\bibitem{Das} A. Das, and C. V. Coffman, J. Math. Phys. {\bf 8}, 1720
(1967).

\bibitem{Boyer} T.H. Boyer, Phys. Rev. {\bf 174}, 1764 (1968).

\bibitem{Wyler} A. Wyler, C.R. Acad. Sci. A{\bf 269}, 743 (1969).

\bibitem{Rosen} G. Rosen, Phys. Rev. D{\bf 13}, 830 (1976).

\bibitem{Chew} G. F. Chew, Phys. Rev. Lett. {\bf 47}, 764 (1981).

\bibitem{Ross} D. K. Ross, Int. J. Theor. Phys. {\bf 25}, 1 (1986).

\bibitem{Dav} P. C. W. Davies, Int. J. Theor. Phys. {\bf 47}, 1949
(2008).

\bibitem{Notealpha0} Note that {\it if} $\alpha$ is identically zero, then
condition (\ref{Eq8}) becomes a trivial one: $2\geq0$. Thus, our
arguments cannot rule out this trivial solution.

\bibitem{Notealpha1} Note that if $\alpha\equiv0$, then
condition (\ref{Eq12}) is satisfied trivially. Thus, our arguments
cannot rule out this trivial solution.

\bibitem{Note1} The other solution of Eq. (\ref{Eq13}) yields the
trivial inequality $\alpha^{1/3}\lesssim140.9$.

\bibitem{Notealpha2} We note again that our arguments cannot rule out the trivial solution $\alpha\equiv0$.
In this case the bound (\ref{Eq1}) is valid trivially.

\end{thebibliography}
\end{document}